\documentclass[3p]{elsarticle}
\pdfoutput=1 
\usepackage{graphicx}
\usepackage{amssymb,amsfonts,amsmath}
\usepackage{tikz}
\usepackage{physics}
\usepackage{bm}
\usepackage{newcommand_yye}
\usepackage{comment}
\usepackage{subcaption}
\usepackage{color}
\usepackage[colorlinks=true,linkcolor=blue,citecolor=red]{hyperref}
\def\ve{\varepsilon}
\def\pa{\partial\Omega}

\def\x{\bm{x}}

\def\pa{{\partial \Omega}}
\def\ve{\varepsilon}

\begin{document}

\begin{frontmatter}
\title{Local persistence exponent and its log-periodic oscillations}

\author[inst1]{Yilin Ye} 
\author[inst1]{Denis~S.~Grebenkov}

\address[inst1]{Laboratoire de Physique de la Mati\`{e}re Condens\'{e}e (UMR 7643), \\ 
CNRS -- Ecole Polytechnique, Institut Polytechnique de Paris, 91120 Palaiseau, France}
\date{Received: \today / Revised version: }

\begin{abstract}
We investigate the local persistence exponent of the survival probability of a particle diffusing near an absorbing self-similar boundary. 
We show by extensive Monte Carlo simulations that the local persistence exponent exhibits log-periodic oscillations over a broad range of timescales.
We determine the period and mean value of these oscillations in a family of Koch snowflakes of different fractal dimensions. 
The effect of the starting point and its local environment on this behavior is analyzed in depth by a simple yet intuitive model. 
This analysis uncovers how spatial self-similarity of the boundary affects the diffusive dynamics and its temporal characteristics in complex systems. 
\end{abstract}


\end{frontmatter} 

\section{Introduction}

The statistics of first-passage times (FPTs) have been extensively studied \cite{Redner, schuss2015brownian, Metzler, Masoliver, Oshanin, Dagdug}. 
In a typical setting, a particle diffuses inside a given Euclidean domain $\Omega$ towards a single or multiple targets, which can be located either inside $\Omega$ or on its reflecting boundary $\pa$. The distribution of the FPT $\mathcal{T}$ to these targets is characterized by the survival probability $\mathbb{P}_{\x_0} \{ \mathcal{T} > t \} = S(t|\x_0)$, where $\x_0$ is the starting point. Many former works were dedicated to the analysis of this distribution, its asymptotic behavior and moments, especially in the case of small targets \cite{Singer06, Condamin07a, Benichou08, Agliari08, Pillay10, Mattos12, Benichou14, Holcman14, Guerin16, Grebenkov16, Simpson21, Balakrishnan19, Meerson15, MejiaMonasterio11, Condamin07}. For instance, when the domain is unbounded, the survival probability 
often exhibits a power-law decay at long times \cite{Bray13, Levernier19}
\begin{equation}
S(t|\x_0) \propto t^{-\alpha} \quad (t \to \infty) \,,
\label{eq:fracdim}
\end{equation} 
where $\alpha$ is called the persistence exponent. 
In this paper, we are interested in a different scenario when the entire boundary $\pa$ is perfectly absorbing, and the survival probability characterizes the ability of a particle started near $\pa$ to avoid absorption as long as possible. 
This setting can also be seen as a zoom into the diffusive dynamics of a particle near an absorbing irregularly-shaped target, and our study aims at understanding how likely for a particle that already approached such a target to move away from it. 

When the boundary is smooth, it can locally be considered as flat. 
In this simplest case, one deals with diffusion in the upper half-plane towards the absorbing horizontal axis, the survival probability is known exactly, and its behavior at long times is characterized by the persistence exponent $\alpha = 1/2$. 
In turn, a fractal boundary can affect the persistence exponent by its irregular geometry \cite{Levitz06, Rozanova12}. 
When particles start uniformly on a contour line at a fixed small distance from the boundary in the plane, the long-time behavior \eqref{eq:fracdim} holds with $\alpha = d_f / 2$, where $d_f$ refers to the fractal (Minkowski) dimension of the boundary. 
In turn, when particles start from a fixed point $\x_0$ near the absorbing self-similar boundary, the long-time behavior of the survival probability $S(t|\x_0)$ turns out to be much richer and can be characterized by 
a \textit{local} persistence exponent (LPE) 
\begin{equation}
\alpha(t|\x_0) = - \frac{\partial \ln S(t|\x_0)}{\partial \ln t} \,. 
\label{eq:alphadef}
\end{equation}
At very short times $t\to0$, $\alpha(t|\x_0)$ goes to 0 due to a finite distance to the boundary; while at very long times, it goes to infinity due to confinement if the domain $\Omega$ is bounded. 
However, in an intermediate time regime, one might expect to get a plateau with a constant value, 
which would thus be identified as the persistence exponent for a limited range of times. In contrast, we have shown in the former study \cite{ye2024first} that, instead of approaching a constant, the LPE exhibit log-periodic oscillations in time. In this situation, the power-law decay \eqref{eq:fracdim} of the survival probability does not hold even at intermediate timescales, and the concept of the persistence exponent appears to be oversimplified and insufficient for describing the much richer diffusion-reaction dynamics.

Log-periodic oscillations present a common feature of self-similar structures.
For instance, one exploited a log-periodic function of the distance to the critical point to describe a critical behavior in the discrete real-space renormalization group \cite{luck2024revisiting}. 
Furthermore, the self-similarity in financial markets may lead to log-periodic oscillations that appear as a potential precursor of crashes \cite{sornette1996stock, Sornette98}.
Among numerous examples of log-periodic oscillations \cite{Akkermans12, Cagnetta15}, we aim at investigating those of the local persistence exponent of the survival probability, related to the ordinary diffusion inside a Euclidean (non-fractal) planar domain with a self-similar boundary, such that $1 < d_f < 2$. 
As our former study \cite{ye2024first} was limited to one example of the Koch snowflake with the vertex angle $\Theta = \frac{\pi}{3}$, and so did another work \cite{ye2025boundary} based on the local time \cite{randon2018residence, Grebenkov20}, our current goal is to refine the log-periodic picture through explorations of LPEs for several families of Koch snowflakes with different vertex angles, orientations, and starting points. 
For this purpose, we perform extensive Monte Carlo simulations in prefractal polygons, which can be considered as finite-scale approximations of a fractal boundary. We inspect the effect of the generation order of such approximations and show that log-periodic oscillations present the intrinsic feature of the fractal boundary. 
We seek to rationalize these oscillations, in particular, to characterize their period and the mean value. 


\section{Local Persistence exponents in Koch snowflakes}
\label{Sec:FPTinK}

\subsection{Koch snowflake}

\begin{figure}[t!]
\centering
\begin{subfigure}{0.3\textwidth}
    \centering
    \includegraphics[height=40mm]{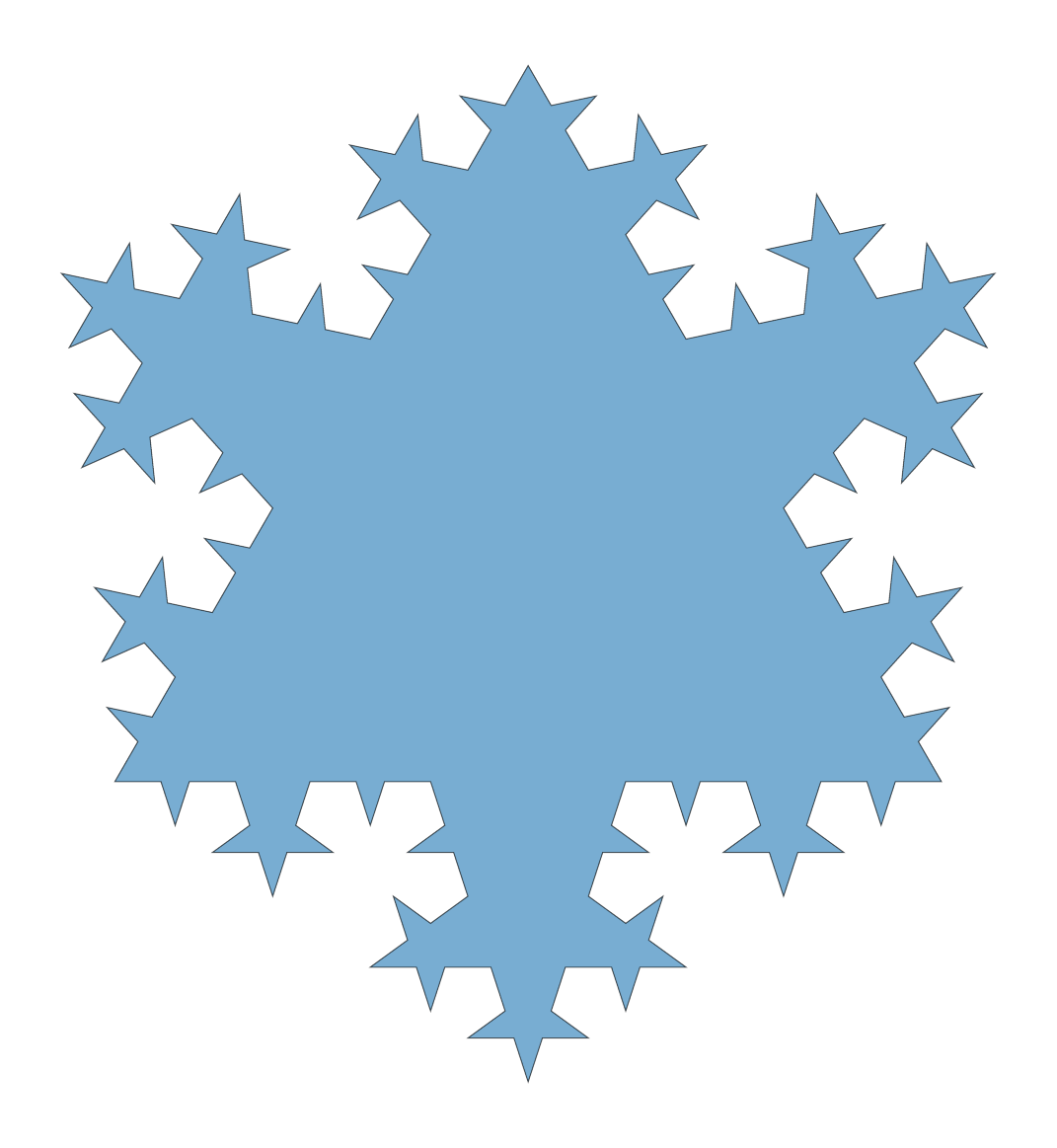}
    \caption{$\Theta = \frac{\pi}{5}$ ($d_f \approx 1.44$). }
    \label{subfig:koch-a}
\end{subfigure}
\hspace{0.2em}
\begin{subfigure}{0.3\textwidth}
    \centering
    \includegraphics[height=40mm]{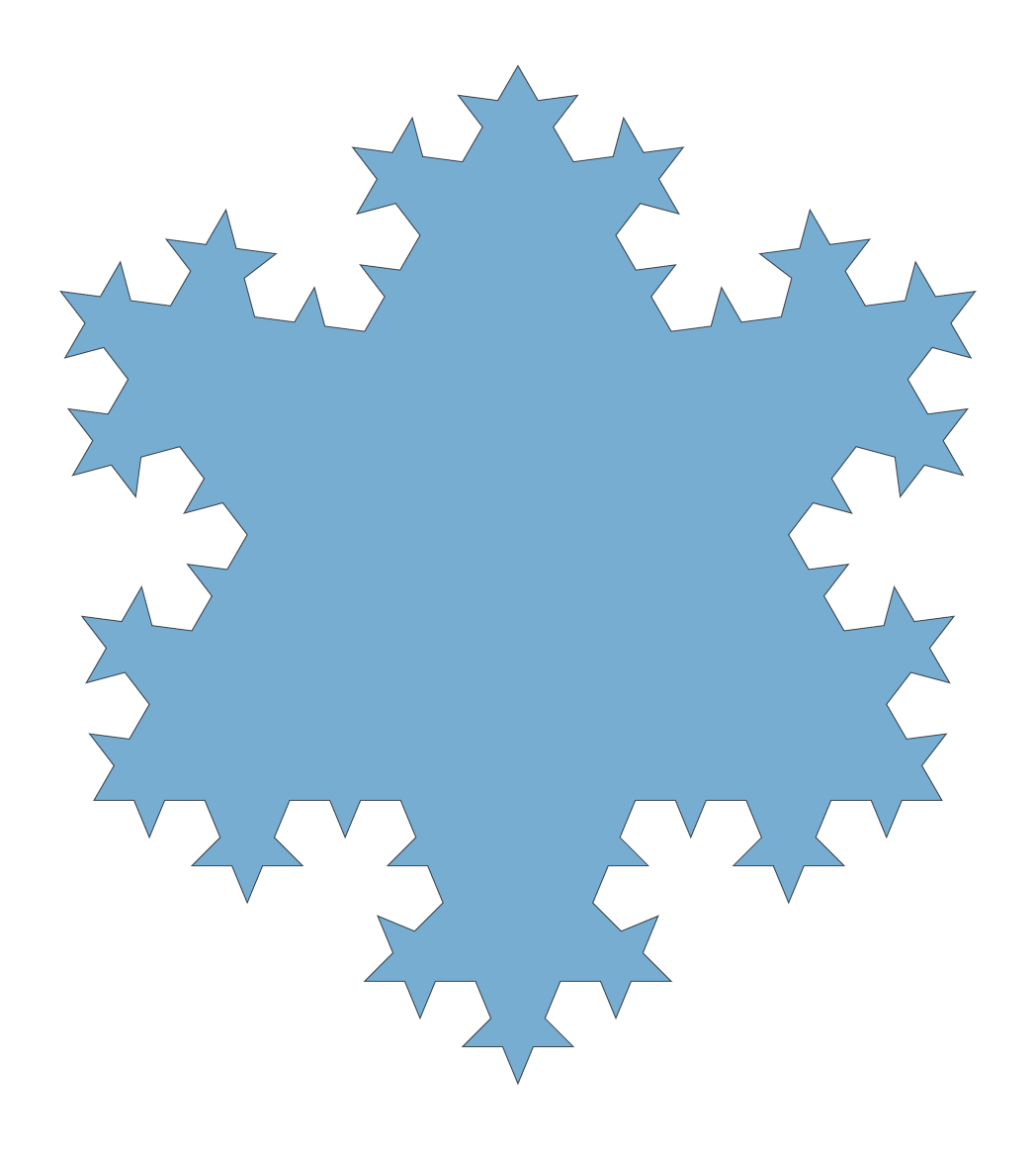}
    \caption{$\Theta = \frac{\pi}{4}$ ($d_f \approx 1.36$). }
    \label{subfig:koch-b}
\end{subfigure}
\hspace{0.2em}
\begin{subfigure}{0.3\textwidth}
    \centering
    \includegraphics[height=40mm]{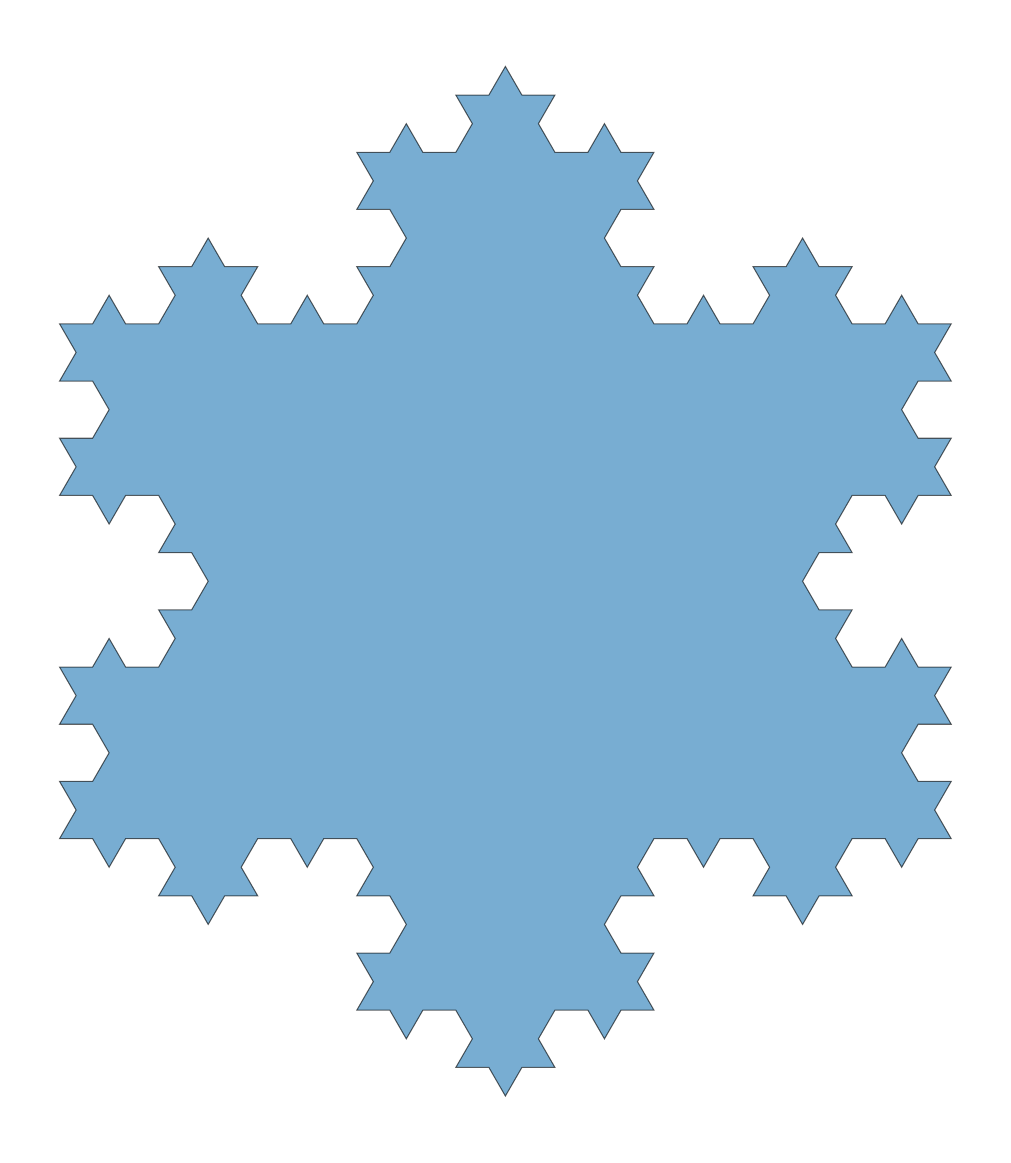}
    \caption{$\Theta = \frac{\pi}{3}$ ($d_f \approx 1.26$). }
    \label{subfig:koch-c}
\end{subfigure}

\vspace{1em} 

\begin{subfigure}{0.3\textwidth}
    \centering
    \includegraphics[height=40mm]{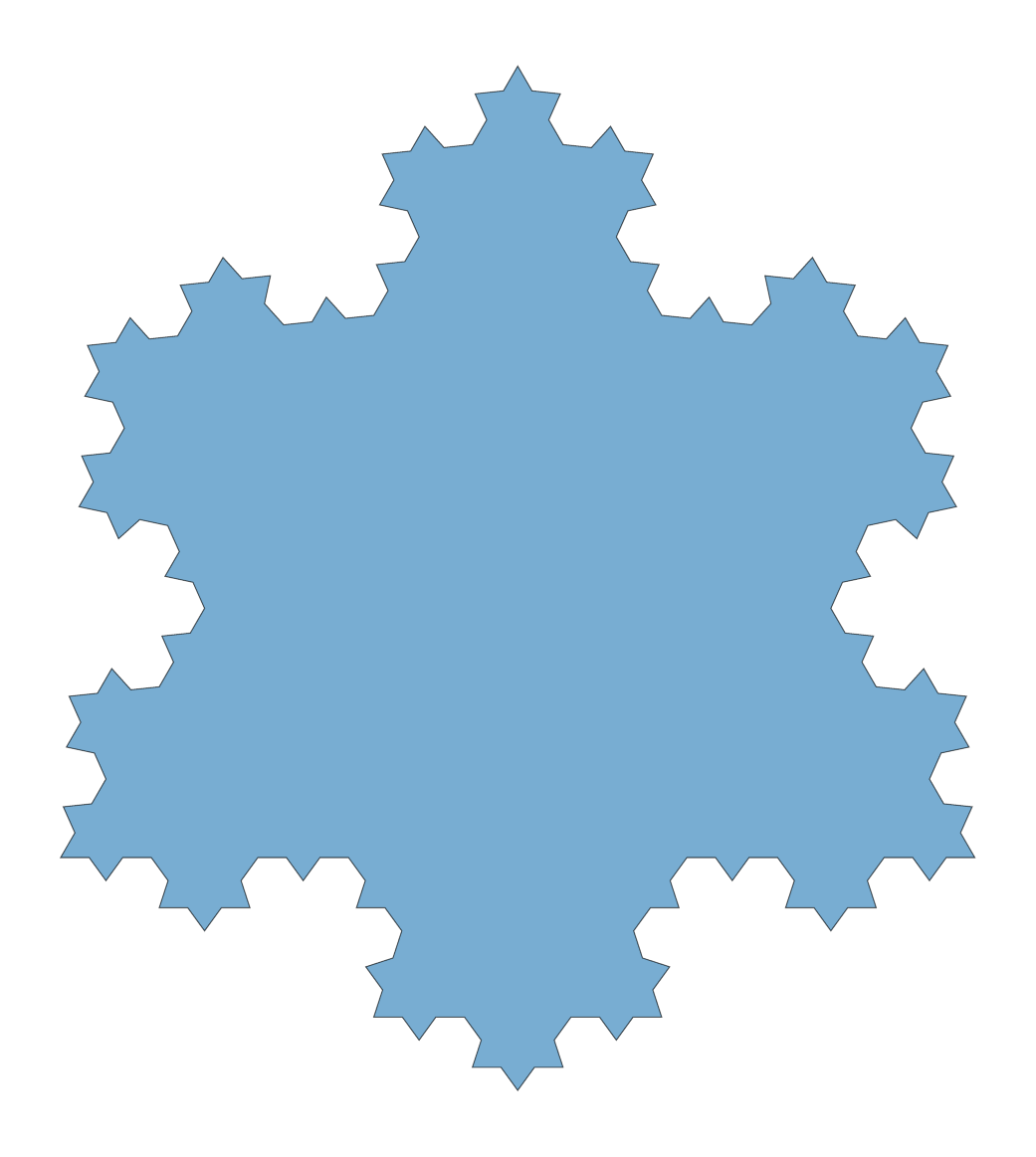}
    \caption{$\Theta = \frac{2\pi}{5}$ ($d_f \approx 1.20$). }
    \label{subfig:koch-d}
\end{subfigure}
\hspace{0.2em}
\begin{subfigure}{0.3\textwidth}
    \centering
    \includegraphics[height=40mm]{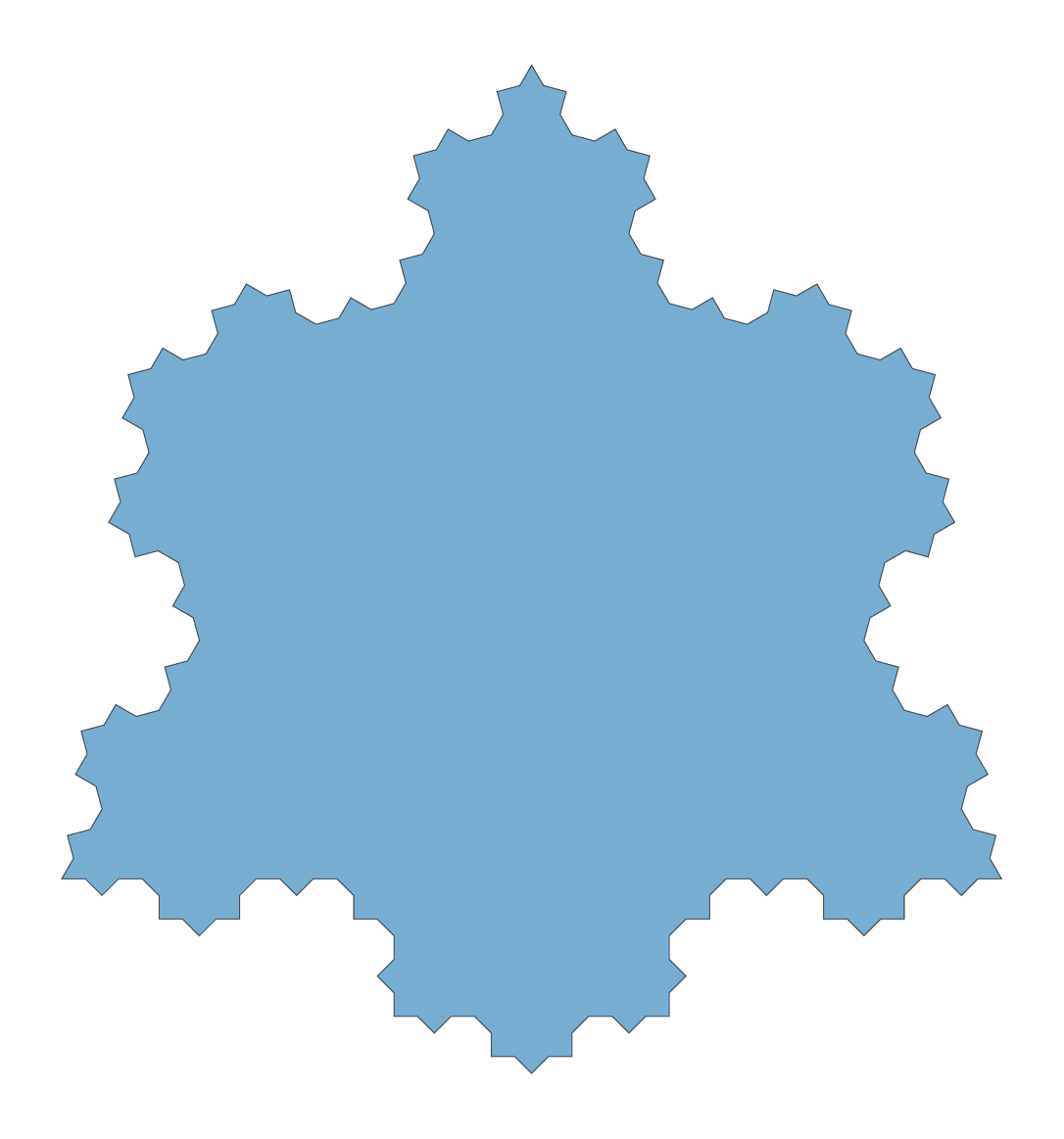}
    \caption{$\Theta = \frac{\pi}{2}$ ($d_f \approx 1.13$). }
    \label{subfig:koch-e}
\end{subfigure}
\hspace{0.2em}
\begin{subfigure}{0.3\textwidth}
    \centering
    \includegraphics[height=40mm]{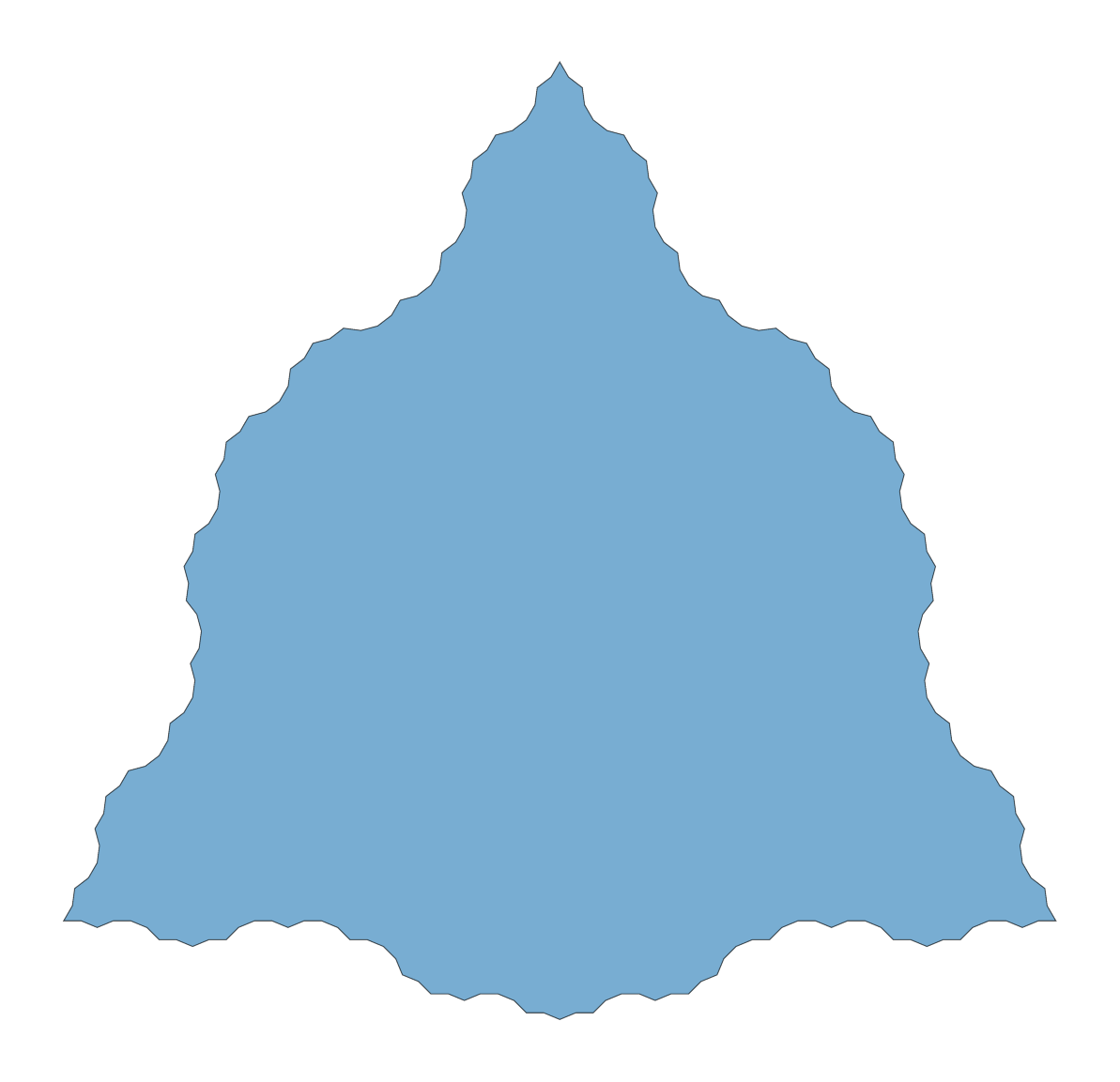}
    \caption{$\Theta = \frac{3\pi}{5}$ ($d_f \approx 1.08$). }
    \label{subfig:koch-f}
\end{subfigure}

\caption{
The third generation $\Omega_{\Theta,3}$ of Koch snowflakes with different angles. 
}
\label{fig:Kochs}
\end{figure}

We consider a family of Koch snowflakes constructed iteratively from a simple generator with a given angle $\Theta$, based on an equilateral triangle $\Omega_0$ with edges of length $L$. 
Each linear segment of the boundary $\partial \Omega_0$ is replaced by a rescaled and appropriately rotated generator to produce the first generation $\Omega_{\Theta,1}$. In the same way, one builds iteratively the successive generations $\Omega_{\Theta,2}$, $\Omega_{\Theta,3}$, $\Omega_{\Theta,4}$, etc. (Fig. \ref{fig:Kochs}). The segments of the boundary $\partial \Omega_{\Theta,g}$ of the $g$-th generation have the length 
\begin{equation}
\ell_{\Theta,g} = \frac{L}{(2[1+\sin(\Theta/2)])^g} \,. 
\label{eq:lg}
\end{equation}
The boundary $\partial \Omega_{\Theta,\infty}$ of the limiting domain $\Omega_{\Theta,\infty}$ is fractal, with the fractal dimension 
\begin{equation}
d_f (\Theta) = \frac{\ln 4}{\ln(2[1+\sin(\Theta/2)])} \,,
\end{equation}
ranging from 1 at $\Theta = \pi$ to 2 as $\Theta \to 0$. 

\begin{figure}[t!]
  \centering
  \begin{subfigure}[b]{0.48\textwidth}
\includegraphics[trim={1.0cm 1.0cm 0.8cm 3.0cm}, clip, width=70mm]{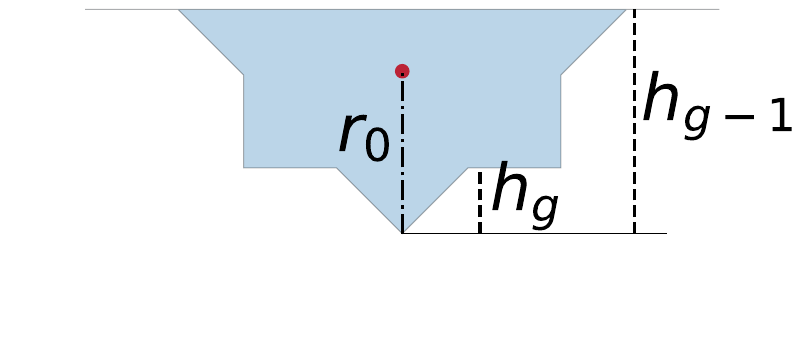}
    \caption{}
    \label{fig:hga}
  \end{subfigure}
  \hspace{1.0em} 
  \begin{subfigure}[b]{0.48\textwidth}
\includegraphics[trim={0.0cm 0.0cm 0.9cm 0.5cm}, clip, height=50mm]{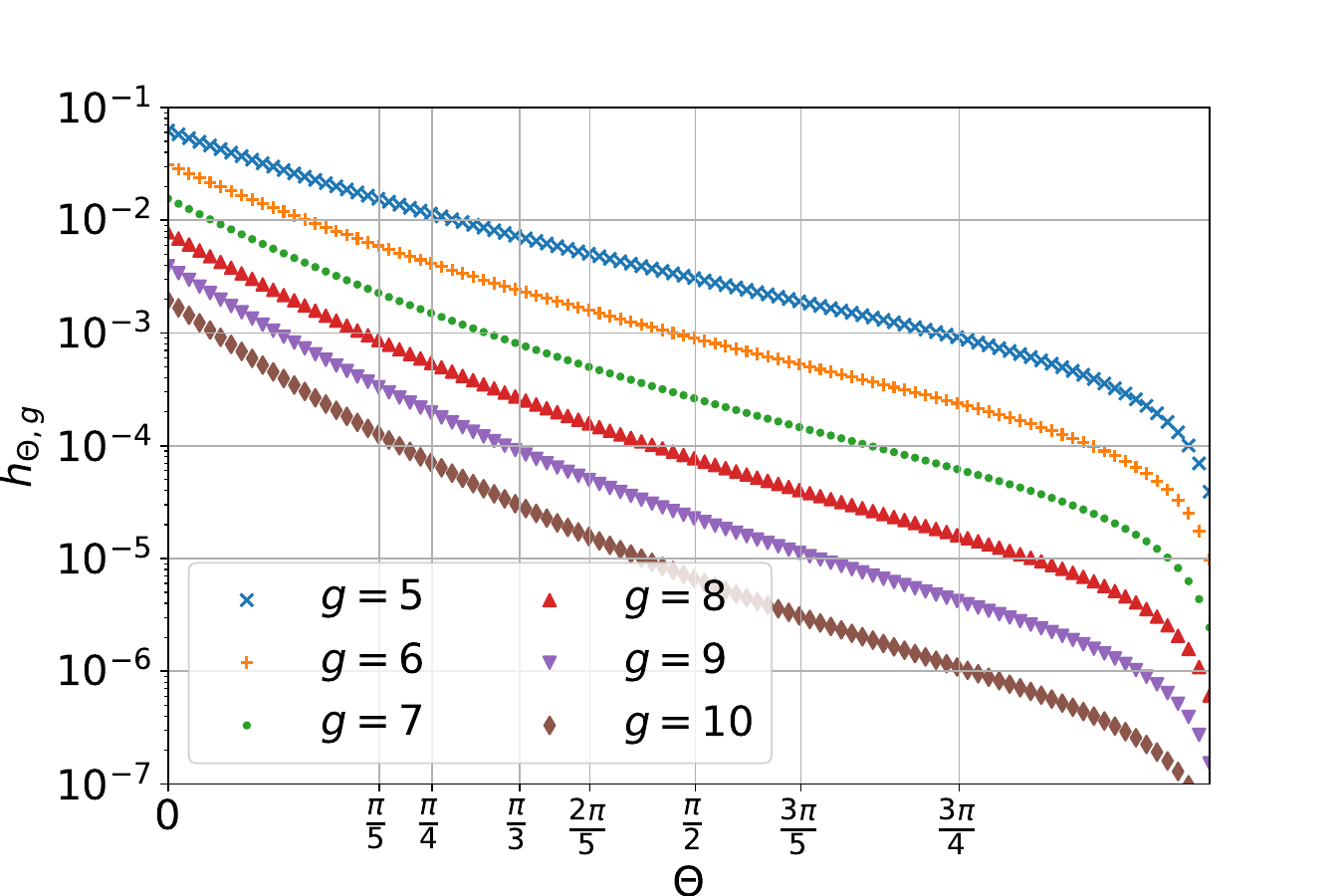}  \caption{}
    \label{fig:hgb}
  \end{subfigure}
  \caption{\textbf{(a)} Local geometry of $\Omega_{\frac{\pi}{2},g}$ with $g>1$. The starting point (in red) is located on the vertical dashed-dotted line close to the bottom vertex with the initial distance $r_0$. 
  The depth $h_g$ of the bottom vertex with respect to the horizontal segment of the $g$-th generation is also shown (for $g$ and $g-1$). 
\textbf{(b)} Depth $h_{\Theta,g}$ as a function of the angle $\Theta$ at different generations $g$ from 5 to 10. }
  \label{fig:hg}
\end{figure}

Throughout this section, 
the starting point is always located on a vertical line close to the bottom vertex of angle $\Theta$, within a vertical distance $r_0$, as illustrated by the red dot on Fig. \ref{fig:hga}. 
As the name suggests, the LPE is highly influenced by the local geometric environment. Thus, its log-periodic oscillation is expected to be controlled by the vertex angle $\Theta$, the generation $g$, and the initial distance $r_0$. 
In the following, we denote the survival probability as $S_{\Theta,g}(t|r_0)$ and the LPE as $\alpha_{\Theta,g}(t|r_0)$.
We emphasize that other choices of the starting point are possible and may lead to different LPEs (see \cite{ye2024first}). As a consequence, our choice (Fig. \ref{fig:hga}) allows us to inspect the effect of the distance $r_0$ in a systematic but not exhaustive way. 

\subsection{Description of Monte Carlo simulations}
\label{sec:desmc}

We resort to Monte Carlo simulations to estimate the survival probability $S_{\Theta,g}(t|r_0)$. 
The geometry-adapted fast random walk (GAFRW) algorithm \cite{Grebenkov05a, Grebenkov05b} is employed for computing rapidly the distance towards the boundary $\partial \Omega_{\Theta,g}$, by exploiting the self-similar structure of Koch snowflakes. 
This is the key step for using the walk-on-spheres (WOS) method \cite{Muller56}.
A random trajectory inside a given domain starts from a fixed point $\x_0$ and continues until the distance to the boundary $\partial \Omega_{\Theta,g}$ is smaller than a prescribed threshold $\varepsilon = 10^{-10}$, which ensures an accurate approximation of the FPT even for $g=10$, because $\ve \ll \ell_{\Theta, 10}$ for all considered angles $\Theta$. 
The duration of a single trajectory is an approximation for the FPT to $\partial \Omega_{\Theta,g}$, while the repetition of this simulations $N$ times provides an empirical statistics of FPTs, allowing one to estimate $S_{\Theta,g}(t|r_0)$. 
For all simulations, we chose $N = 10^8$ to ensure a sufficient statistics. 
To achieve more accurate statistics of rare events of large FPTs, we executed additional simulations by introducing a prescribed time threshold $T_\mathrm{min}$, such that only the generated FPTs above $T_\mathrm{min}$ were recorded to accumulate $N = 10^8$ statistics. 
A home-built code was written in Fortran 90. Numerous datasets of FPTs were analyzed in MATLAB by the standard routine ``ecdf'' for obtaining the empirical cumulative distribution function (CDF). 
To estimate the LPE from Eq. (\ref{eq:alphadef}), we define a range of $K$ times $t_k = t_0 (t_K/t_0)^{k/K}$ (with $k = 1, \ldots, K$), which are equally spaced on the logarithmic scale between prescribed bounds $t_0$ and $t_K$. To interpolate $\alpha_{\Theta,g}(t|r_0)$ from the empirical CDF, we set
\begin{equation} 
\alpha_{\Theta,g}(t'_k|r_0) = - \frac{\ln(S_{\Theta,g}(t_{k}|r_0)/S_{\Theta,g}(t_{k-1}|r_0))}{\ln (t_{k}/t_{k-1})}, 
\label{eq:alphainterpret}
\end{equation} 
estimated at the intermediate time $t'_k = \sqrt{t_k t_{k-1}}$ between $t_{k-1}$ and $t_k$ (the geometric mean is used to respect the logarithmic spacing of points).

Throughout the paper, we fix time and length scales by setting $L = 2$ and $D = 1$. 
To analyze the log-periodic behavior of LPEs, we set $K =400$, $t_0 \lesssim r_0^2 \sin^2(\Theta/2)/(4D)$, and $t_K \gtrsim 10^6 \, t_0 $. We manually extract the coordinates $\{(T_{\Theta, i}^\pm, \alpha_{\Theta, i}^\pm)\}$ of each maximum (plus sign) and minimum (minus sign) from the dataset given by Eq. \eqref{eq:alphainterpret}, which is checked systematically by the script ``scipy.signal.argrelextrema'' in Python. 
The periods, mean values, and amplitudes of oscillations of $\alpha_{\Theta,g}(t|r_0)$ are then derived from these coordinates.

\subsection{Local persistence exponents}

In the previous work \cite{ye2024first}, the LPE was investigated only for the conventional Koch snowflake with $\Theta = \frac{\pi}{3}$. 
How does the angle $\Theta$ and thus the fractal dimension affect log-oscillations? 
Before presenting the results, we recall three distinct regimes identified in \cite{ye2024first}: 
(i) at short times, $t \ll r_0^2/D$, 
the survival probability behaves as $1 - S_{\Theta,g}(t|r_0) \propto e^{-\delta^2 / (4Dt)}$ where $\delta \propto r_0$ is the distance to the boundary, and thus $\alpha_{\Theta,g}(t|r_0) \propto e^{-\delta^2/(4Dt)}$ grows monotonously from 0 as $t$ increases; 
(ii) at intermediate times, $r_0^2 / D \lesssim t \ll L^2/D$, the particle diffuses near a self-similar boundary, and $\alpha_{\Theta,g}(t|r_0)$ is expected to exhibit a plateau or an oscillation; 
(iii) at long times, $t \gtrsim L^2/D$, the particle leaves the self-similar region and enters the central part of the confining domain, so that $S_{\Theta,g}(t|r_0) \propto e^{-D \lambda_1 t}$, where $\lambda_1$ is the principal eigenvalue of the Dirichlet Laplacian, and thus $\alpha_{\Theta,g}(t|r_0) \propto t$ as $t\to\infty$. 
We are primarily interested in the intermediate, least understood regime. 
Even though the effect of the timescales $r_0^2/D$ and $L^2/D$ can be observed in plots of $\alpha_{\Theta,g}(t|r_0)$ (see below), we propose a more direct, semi-quantitative way to estimate coarsely the number of oscillations according to the starting distance $r_0$.

\begin{figure}[t!]
  \centering
  \begin{subfigure}[b]{0.48\textwidth}
\includegraphics[trim={0.2cm 0.1cm 1.0cm 1.0cm}, clip, width=0.99\linewidth]{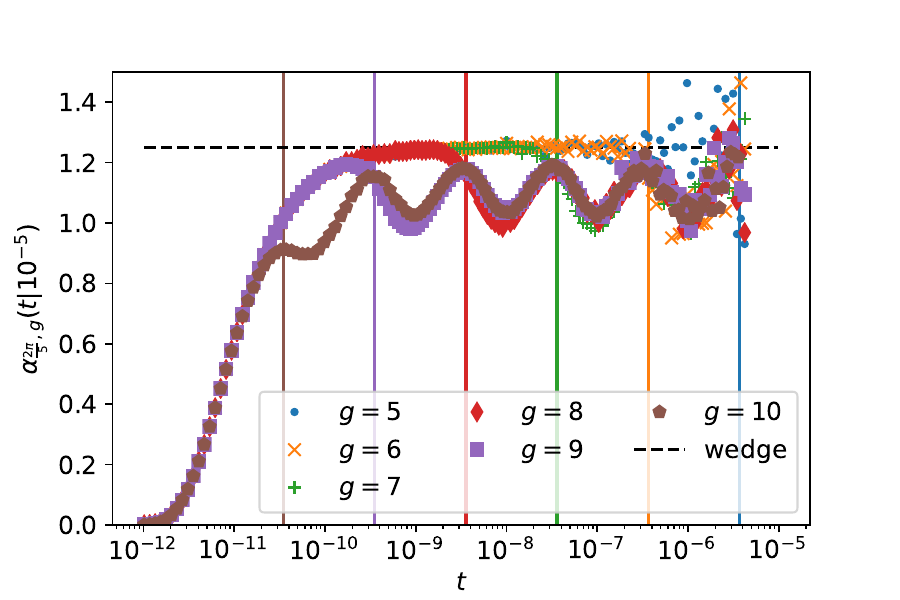}
    \caption{}
    \label{fig:2pi5gt-a}
  \end{subfigure}
  \hspace{1.0em} 
  \begin{subfigure}[b]{0.48\textwidth}
\includegraphics[trim={0.3cm 0.1cm 1.0cm 1.0cm}, clip, width=0.99\linewidth]{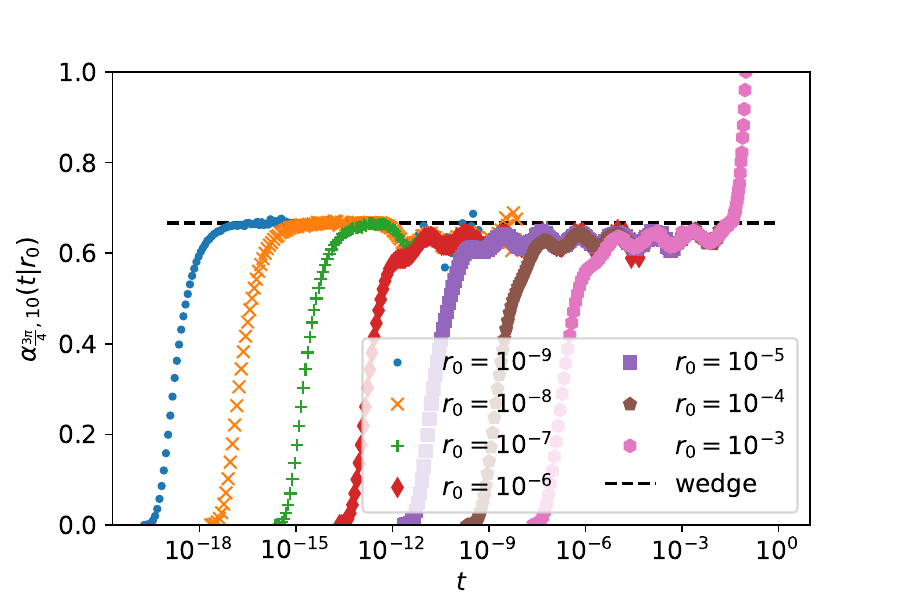}
  \caption{}
    \label{fig:2pi5gt-b}
  \end{subfigure}
  \caption{\textbf{(a)} Local persistence exponent $\alpha_{\frac{2\pi}{5},g}(t|10^{-5})$ for six generations (from 5 to 10), with $L = 2$, and $D = 1$, obtained by Monte Carlo simulations with $N = 10^8$ particles. The time threshold $T_\mrm{min} = 6\times10^{-10}$ is taken for $g$ from 5 to 9, and $T_\mrm{min} = 8\times10^{-10}$ for $g=10$. Dashed horizontal line presents the persistence exponent $\frac{5}{4}$ for the wedge of angle $\frac{2\pi}{5}$. Vertical lines present $h_{\frac{2\pi}{5},g}^2/D$ for each $g$ in the corresponding color. 
\textbf{(b)} Local persistence exponent $\alpha_{\frac{3\pi}{4},10} (t|r_0)$ for different starting points $r_0$, with $L = 2$, and $D = 1$, obtained by Monte Carlo simulations with $N = 10^8$ particles. The time threshold $T_\mrm{min} = 500 r_0^2 / D$ is taken. Dashed horizontal line presents the persistence exponent $\frac{2}{3}$ for the wedge of angle $\frac{3\pi}{4}$. }
  \label{fig:2pi5gt}
\end{figure}

To characterize the role of the initial distance, we introduce the depth $h_{\Theta,g}$, i.e., the distance between the bottom vertex and the closest horizontal edge of $\Omega_{\Theta,g}$ (Fig. \ref{fig:hga}): 
\begin{equation}
h_{\Theta,g} = \ell_{\Theta,g} \cos(\Theta/2) = \frac{L \cos(\Theta/2)}{(2[1+\sin(\Theta/2)])^g} \,. 
\label{eq:hg}
\end{equation}
Figure \ref{fig:hgb} shows the depth $h_{\Theta,g}$ as a function of angle $\Theta$ as $g$ goes from 5 to 10, which helps us to choose a suitable $r_0$ for each $\Theta$ and $g$ by comparing it to $h_{\Theta,g}$. 
When $r_0 \ll h_{\Theta,g}$, the particle starts too close to the vertex and thus first explores a triangular (or wedge-looking) domain. 
As a consequence, the survival probability $S_{\Theta,g}(t|r_0)$ at small-intermediate times is expected to be close to the survival probability in a wedge of angle $\Theta$, which exhibits the power-law decay with the persistence exponent $\alpha = \frac{\pi}{2\Theta}$ \cite{Redner, Considine89, Comtet03, Dy08, Chupeau15, LeVot20}. 
In other words, one can expect a plateau of the LPE at small-intermediate timescales. 
In turn, if $r_0 \gtrsim h_{\Theta,g}$, the particle explores the prefractal domain, and the associated LPE may exhibit log-oscillations. Let us illustrate these points. 

Figure \ref{fig:2pi5gt-a} presents the LPE $\alpha_{\frac{2\pi}{5},g} (t|r_0)$ with $g$ from 5 to 10, and $r_0 = 10^{-5}$. 
At short times ($t \ll r_0^2/D$), all curves show the same monotonous increase. 
Then, at $t\approx2\times10^{-11}$, the curve of $\alpha_{\frac{2\pi}{5},10}(t|r_0)$ deviates from the others and starts oscillating. 
Since $r_0 \sim h_{\frac{2\pi}{5},10} \approx 1.55 \times 10^{-5}$, the particle is located in a prefractal local environment of the generation $g=10$, thus resulting in the first peak at $t \approx 3\times10^{-11}$ of a lower amplitude and the next peak situated around $t \approx 3\times 10^{-10}$. 
In turn, $\alpha_{\frac{2\pi}{5},9}(t|r_0)$ deviates from the other curves at $t \approx 2 \times 10^{-10}$ and rejoins $\alpha_{\frac{2\pi}{5},10}(t|r_0)$. 
For $t \gtrsim 3\times10^{-9}$, both curves show the same log-oscillations.  
For the 8-th generation, one has $r_0 \ll h_{\frac{2\pi}{5},8}$, so that the particle ``feels'' a wedge for a longer timespan, so that the curve $\alpha_{\frac{2\pi}{5},8}(t|r_0)$ has enough time to reach the plateau $\frac{\pi}{2\Theta} = \frac{5}{4}$, before entering the oscillation regime. 
One sees that the relation between $r_0$ and $h_{\Theta,g}$ (Fig. \ref{fig:hgb}) helps to determine whether $\alpha_{\Theta,g}(t|r_0)$ would reach the wedge plateau or not. 
The LPE $\alpha_{\frac{2\pi}{5},8}(t|r_0)$ steps into the oscillation regime with one-period delay as compared to $\alpha_{\frac{2\pi}{5},9}(t|r_0)$, and so recursively for smaller $g$. 
A faster decay of the survival probability $S_{\frac{2\pi}{5},5}(t|r_0)$ made more difficult an accurate estimation of the LPE $\alpha_{\frac{2\pi}{5},5}(t|r_0)$ at long times, even by using a time threshold $T_\mrm{min} = 8\times10^{-10}$. 
As a consequence, the noisy scattered values of this LPE at long times do not allow us to observe the expected oscillation regime for $t$ in the range between $3\times 10^{-7}$ and $3\times 10^{-6}$. 
In turn, for all generations from 6 to 10, we conclude that their LPEs $\alpha_{\frac{2\pi}{5},g}(t|r_0)$ converge to the \tit{same} oscillation pattern, which can thus be qualified as the intrinsic feature of the fractal boundary. In fact, the generation $g$ only affects the time instance when $\alpha_{\frac{2\pi}{5},g}(t|r_0)$ reaches this pattern, with larger $g$ corresponding to earlier times. 
In other words, even though it is not possible to run Monte Carlo simulations for the truly fractal boundary ($g = \infty$), finite generations of the fractal give access to the asymptotic behavior of the survival probability $S_{\Theta,\infty}(t|r_0)$ of the fractal boundary $\partial\Omega_{\Theta,\infty}$. 
As shown by colored vertical lines in Fig. \ref{fig:2pi5gt-a}, the time instances when different curves start oscillations can be estimated by $h_{\frac{2\pi}{5},g}^2/D$.

In addition to the influence of the generation $g$, the effect of the initial distance $r_0$ is explored by focusing on the LPE $\alpha_{\frac{3\pi}{4},10}(t|r_0)$ for generation $g = 10$ (Fig. \ref{fig:2pi5gt-b}), when $r_0$ ranges from $10^{-9}$ to $10^{-3}$ (note that $h_{\frac{3\pi}{4},10} \approx 1.08\times10^{-6}$). 
We recall that, if $r_0 \ll h_{\frac{3\pi}{4},10}$ is satisfied, $\alpha_{\frac{3\pi}{4},10}(t|r_0)$ reaches the wedge plateau $\frac{\pi}{2\Theta} = \frac{2}{3}$ and stays there, before switching to log-periodic oscillations. 
This is clearly seen for $r_0 = 10^{-9}, 10^{-8}, 10^{-7}$. 
For all other distances satisfying $r_0 \gtrsim h_{\frac{3\pi}{4},10}$, the LPE $\alpha_{\frac{3\pi}{4},10}(t|r_0)$ does not reach the wedge plateau and starts log-periodic oscillations directly after a transient growth. 
Moreover, for the largest distance $r_0 = 10^{-3}$, we observe the ultimate confinement regime $\alpha_{\frac{3\pi}{4},10}(t|r_0) \propto t$ at long times when the constraint $t \ll L^2 / D$ is not satisfied. 
Importantly, all curves in the log-oscillation regime fall on each other, due to the self-similarity of the boundary. 
{This indicates independence of the LPE behavior on $r_0$ once $g$ is large enough to ensure that $r_0 \gtrsim h_{\Theta,g}$.}
To observe at least one period of oscillations in the intermediate timescale before the linear growth, one needs to set the starting point with $r_0 \ll h_{\Theta,1}$. 
Before reaching the threshold $h_{\Theta,1}$, the particle ``feels'' a self-similar boundary, leading to the same log-periodic oscillation in the intermediate time region (by mean value and amplitude) for different $g$. 
In turn, for a starting point satisfying $r_0 \gtrsim h_{\Theta,1}$, or for a long time $t \gtrsim L^2/D$, the particle is already at the central region of the equilateral triangle $\Omega_0$ and thus far away from the self-similary boundary, that prohibits log-periodic oscillations.  

\begin{figure}[t!]
  \centering
  \begin{subfigure}[b]{0.48\textwidth}
\includegraphics[trim={0.4cm 0.5cm 0.5cm 0.3cm}, clip, width=0.99\linewidth]{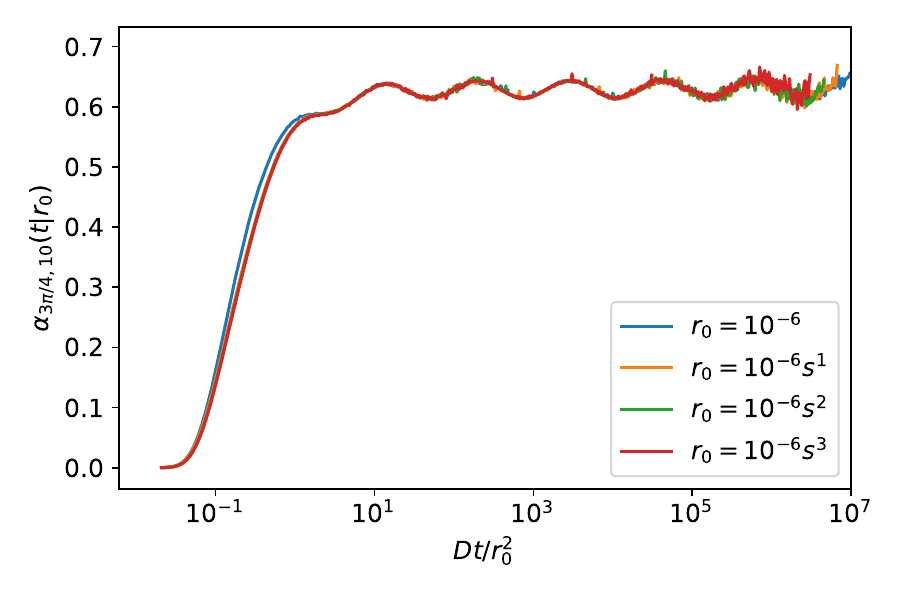}
    \caption{}
    \label{fig:3pi4gt-a}
  \end{subfigure}
  \hspace{1.0em} 
  \begin{subfigure}[b]{0.48\textwidth}
\includegraphics[trim={0.4cm 0.5cm 0.5cm 0.3cm}, clip, width=0.99\linewidth]{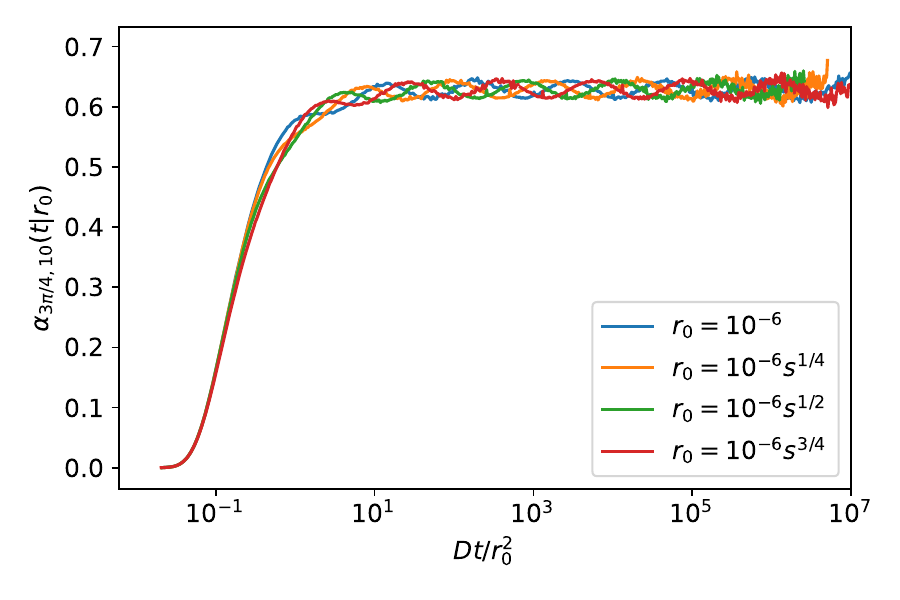}
  \caption{}
    \label{fig:3pi4gt-b}
  \end{subfigure}
\caption{
\textbf{(a)} Local persistence exponent $\alpha_{\frac{3\pi}{4},10}(t|r_0)$ as a function of $\tilde{t} = Dt/r_0^2$ for different starting points $r_0 = 10^{-6} s^{i}$ ($i$ from 0 to 3), with $L = 2$, $D = 1$, and $s = 2(1+\sin(\Theta/2)) \approx 3.84$, obtained by Monte Carlo simulations with $N = 10^8$ particles. The time threshold $T_\mrm{min} = 500 r_0^2 / D$ is taken. 
\textbf{(b)} Local persistence exponent $\alpha_{\frac{3\pi}{4},10}(t|r_0)$ as a function of $\tilde{t}$ for different starting points $r_0 = 10^{-6} s^{i/4}$ ($i$ from 0 to 3), with $L = 2$, and $D = 1$, obtained by Monte Carlo simulations with $N = 10^8$ particles. The time threshold $T_\mrm{min} = 5\times10^{-10}$ is taken for $i=0,1,2$ and $T_\mrm{min} = 5\times10^{-9}$ for $i=3$.
}
  \label{fig:3pi4gt}
\end{figure}

Moreover, a further inspection of Fig. \ref{fig:2pi5gt-b} reveals the LPE curves exhibit the same short-time transient growth, which is just shifted in time for different $r_0$. 
To validate this observation, one can perform an appropriate horizontal shift by using the rescaled time $\tilde{t} = Dt/r_0^2$. 
However, such a time rescaling does not necessarily correct for eventual phase differences that may exist among oscillations for different values of $r_0$. 
To account for such phase differences, we introduce the scaling factor  $s = h_{\Theta,g-1} / h_{\Theta,g} = 2 (1+\sin(\Theta/2))$ and then choose the initial distances $r_0$ accordingly. 
Figure \ref{fig:3pi4gt-a} presents the LPE $\alpha_{\frac{3\pi}{4},g}(t|r_0)$ as a function of $\tilde{t}$ for $r_0 = 10^{-6} s^i$ with $i$ from 0 to 3. 
Remarkably, all these LPE curves collapse onto each other, which illustrates how the spatial self-similarity of the boundary affects the diffusive dynamics via log-oscillations. 
At short times, there is a little deviation in the transient growth, because $h_{\frac{3\pi}{4},10} \approx 1.08 \times 10^{-6}$ is comparable to the smallest $r_0 = 10^{-6}$. 
As zero phase difference in oscillations corresponds to integer powers of $s$, non-integer powers may be used to control phase shifts. 
As shown in Fig. \ref{fig:3pi4gt-b}, the curves of the LPE $\alpha_{\frac{3\pi}{4},g}(t|r_0)$ with $r_0 = 10^{-6} s^{i/4}$ (with $i$ from 0 to 3) exhibit a clear $\frac{\pi}{2}$ phase difference between two adjacent curves. 


\subsection{Period of log-oscillations}
\label{sec:logperiod}

\begin{figure}[t!]
  \centering
  \begin{subfigure}[b]{0.48\textwidth}
\includegraphics[trim={0.5cm 0.8cm 0.6cm 0.3cm}, clip, width=0.99\linewidth]{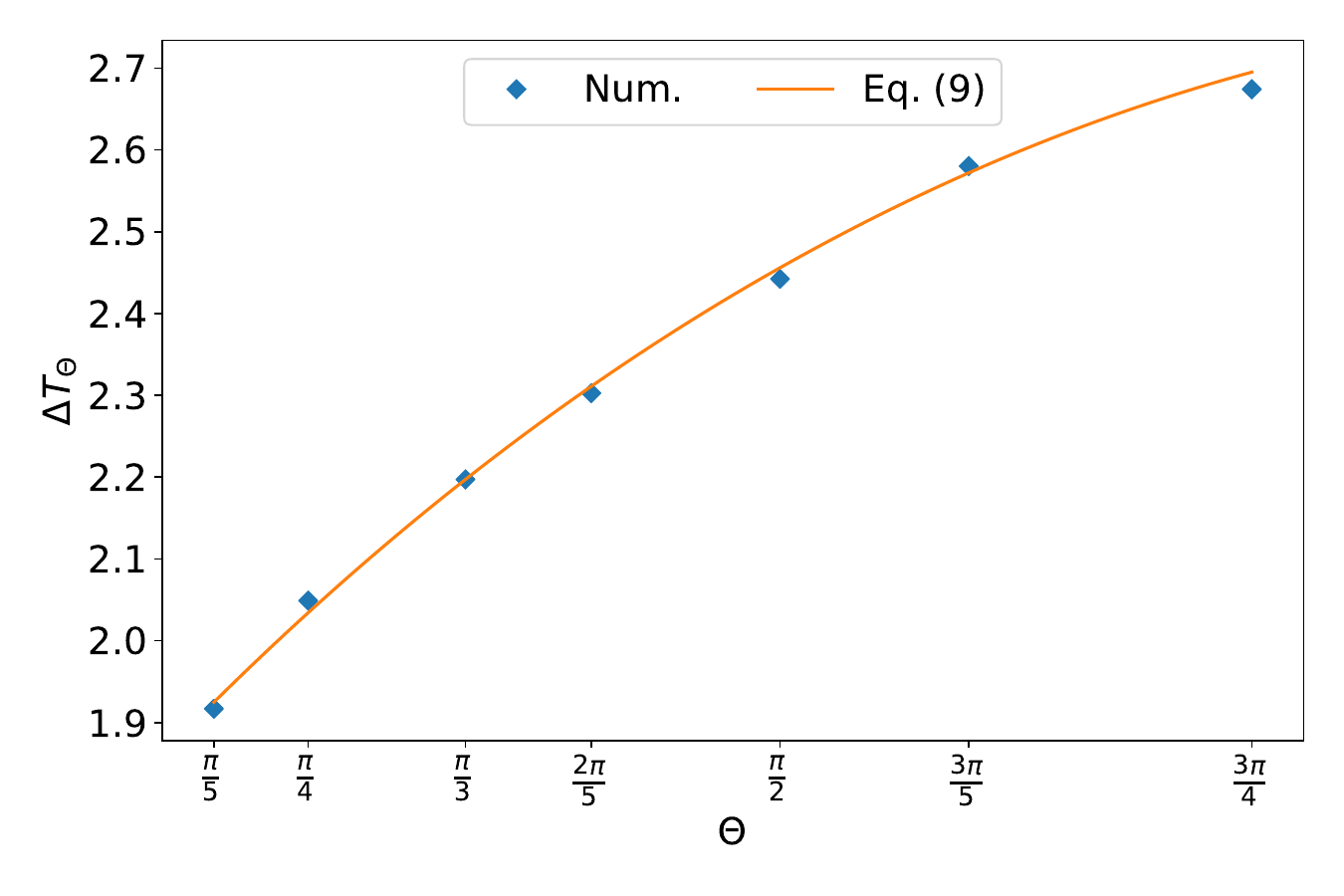}
    \caption{}
    \label{fig:vTlog}
  \end{subfigure}
  \hspace{1.0em} 
  \begin{subfigure}[b]{0.48\textwidth}
\includegraphics[trim={0.2cm 0.8cm 0.6cm 0.3cm}, clip, width=0.99\linewidth]{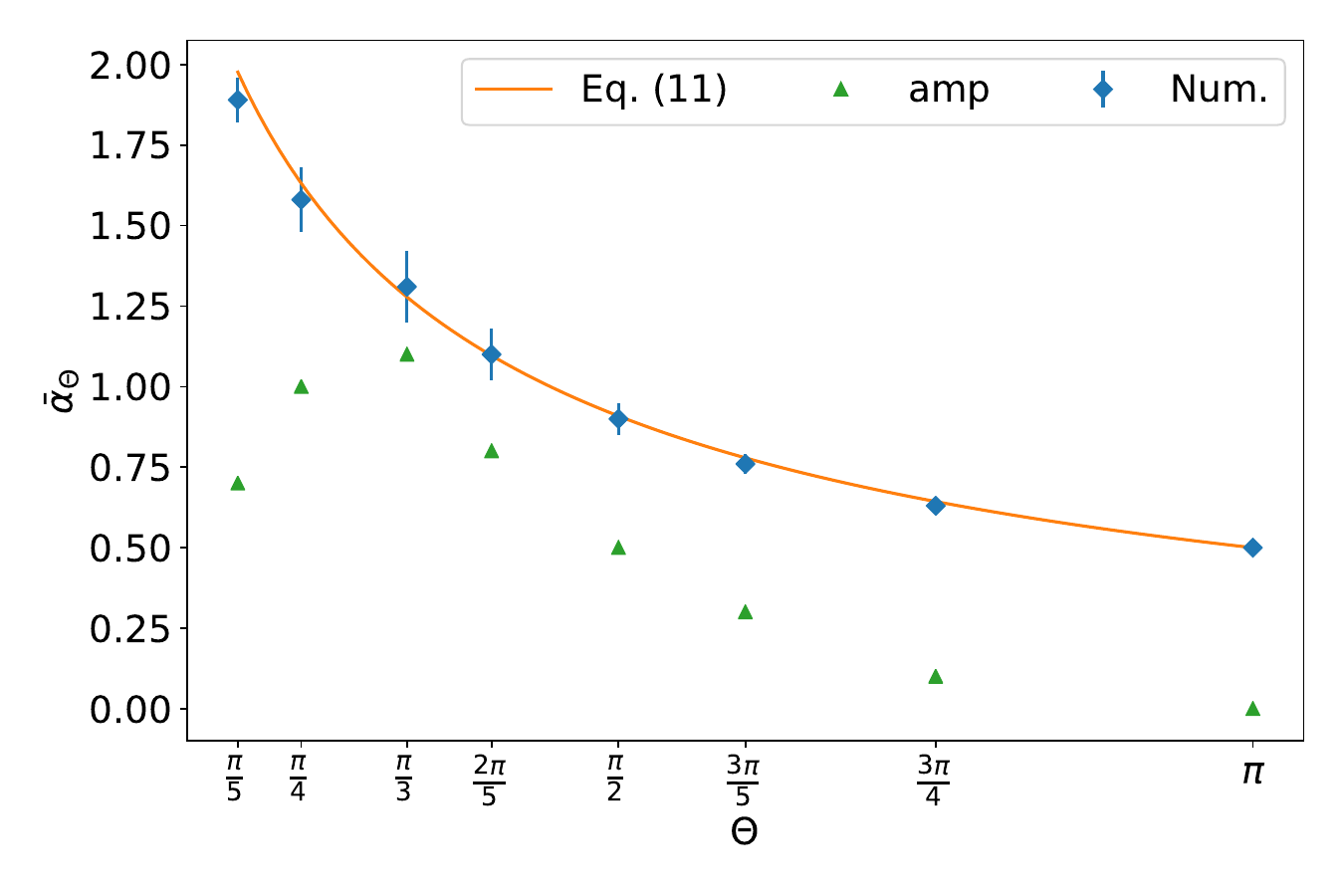}
  \caption{}
    \label{fig:amv}
  \end{subfigure}
  \caption{
  \textbf{(a)} Comparison of log-periods $\Delta T_\Theta$ between numerical results (diamonds) and analytical approximation (solid line) by Eq. \eqref{eq:validTlog}. Angles are selected as $\Theta = \frac{\pi}{5}$, $\frac{\pi}{4}$, $\frac{\pi}{3}$, $\frac{2\pi}{5}$, $\frac{\pi}{2}$, $\frac{3\pi}{5}$, $\frac{3\pi}{4}$. 
\textbf{(b)} Comparison of mean LPEs between numerical results (diamonds) and $\bar{\alpha}^\mrm{eff}_{\Theta,g}$ (solid line) from Eq. (\ref{eq:meanalpha}). 
The last point $\Theta = \pi$ is added manually with $\alpha_\pi = 
\frac{1}{2}$ for a flat boundary. 
The amplitude $\Delta\alpha_{\Theta}$ of oscillations is shown by vertical errorbars and is also presented separately by triangles that show $10 \Delta \alpha_\Theta$ for a better visualization. 
}
\end{figure}

The examples shown in Fig. \ref{fig:2pi5gt} suggest the presence of log-periodic oscillations of LPEs $\alpha_{\Theta,g}(t|r_0)$ for various $\Theta$ and $g$ in the intermediate time regime. 
As a function of angle $\Theta$, we denote $\Delta T_\Theta$ the log-period of $\alpha_{\Theta,g}(t|r_0)$. 
As illustrated above, the parameters $g$ and $r_0$ determine the range of log-periodic oscillations but do not affect their log-period; in other words, $\Delta T_\Theta$ is indeed a function of $\Theta$ (or $d_f$, the fractal dimension) alone. 
By expressing the time instances of $i$-th maximum and minimum on the $\alpha_{\Theta,g}(t|r_0)$ curve as $T^{+}_{\Theta,i}$ and $T^{-}_{\Theta,i}$, the log-period reads 
\begin{equation} 
\Delta T_\Theta = \log T^{+}_{\Theta,i+1} - \log T^{+}_{\Theta,i} = \log T^{-}_{\Theta,i+1} - \log T^{-}_{\Theta,i} \,, 
\end{equation} 
which is supposed to be independent of $i$. 
This log-period demonstrates a geometric regression in time, and we need to figure out its common ratio $\lambda_\Theta$ as a function of $\Theta$, i.e.,  $\lambda_\Theta = T^\pm_{\Theta,i+1} / T^\pm_{\Theta, i} = \exp \llp \Delta T_\Theta \rrp$. 
In order to relate the log-period $\Delta T_\Theta$ to the geometric structure of the local environment, we analyze a typical area ``covered'' during diffusion up to time $t$. 
We denote $A_{\Theta,g}$ the area of one small triangle generated on the segment $\ell_{\Theta,g-1}$ when passing from $\Omega_{\Theta,g-1}$ to $\Omega_{\Theta,g}$: 
\begin{equation}
A_{\Theta,g} = \ell_{\Theta,g}^2 \sin \frac{\Theta}{2} \cos \frac{\Theta}{2} = \frac{L^2 \sin\Theta}{2(2[1+\sin(\Theta/2)])^{2g}} \,.
\label{eq:Aareag}
\end{equation}
As the log-periodic oscillations originate from the self-similar structure of the boundary, we conjecture that the log-period depends on the ratio of areas in two successive generations as 
\begin{equation}
\Delta T_\Theta \approx \ln \frac{A_{\Theta,g-1}}{A_{\Theta,g}} = 2\ln \lls 2 \llp 1 + \sin \frac{\Theta}{2} \rrp \rrs = \frac{4\ln2}{d_f} \,.
\label{eq:validTlog}
\end{equation}
Figure \ref{fig:vTlog} compares the numerical results obtained from various curves of the LPEs $\alpha_{\Theta,g}(t|r_0)$ to the approximation (\ref{eq:validTlog}). 
One sees an excellent agreement, despite the simplicity of the underlying argument. 
Indeed, there is no obvious tendency of relative errors and their absolute values are below $1\%$, justifying the approximation (\ref{eq:validTlog}). 

\subsection{Mean value of $\alpha_{\Theta,g}(t|r_0)$}
\label{sec:meanalpha}

\begin{figure}[t!]
  \centering
  \begin{subfigure}[b]{0.48\textwidth}
\includegraphics[trim={1.5cm 2.0cm 0.5cm 1.5cm}, clip, width=0.99\linewidth]{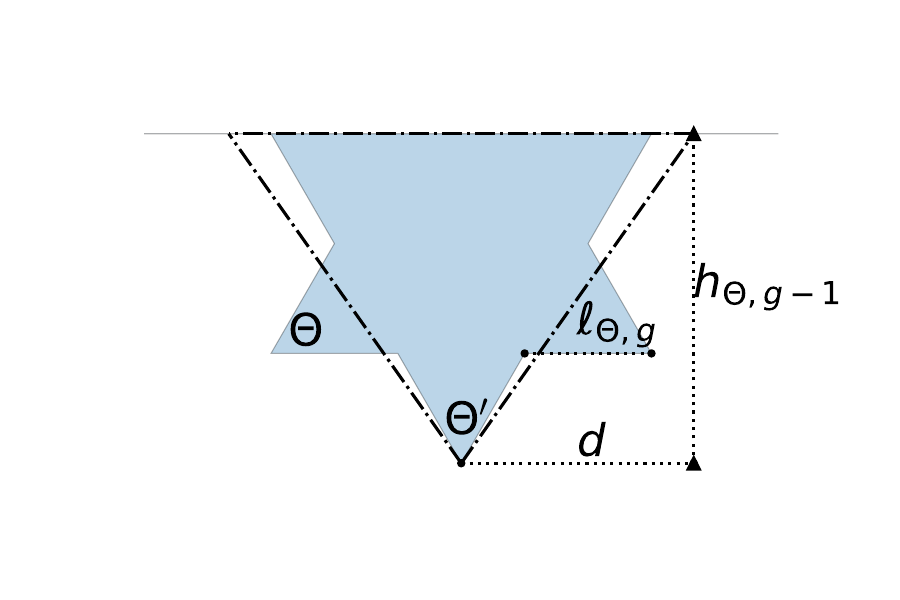}
    \caption{}
    \label{fig:trimod}
  \end{subfigure}
  \hspace{1.0em} 
  \begin{subfigure}[b]{0.48\textwidth}
    \centering
\includegraphics[width=0.5\linewidth]{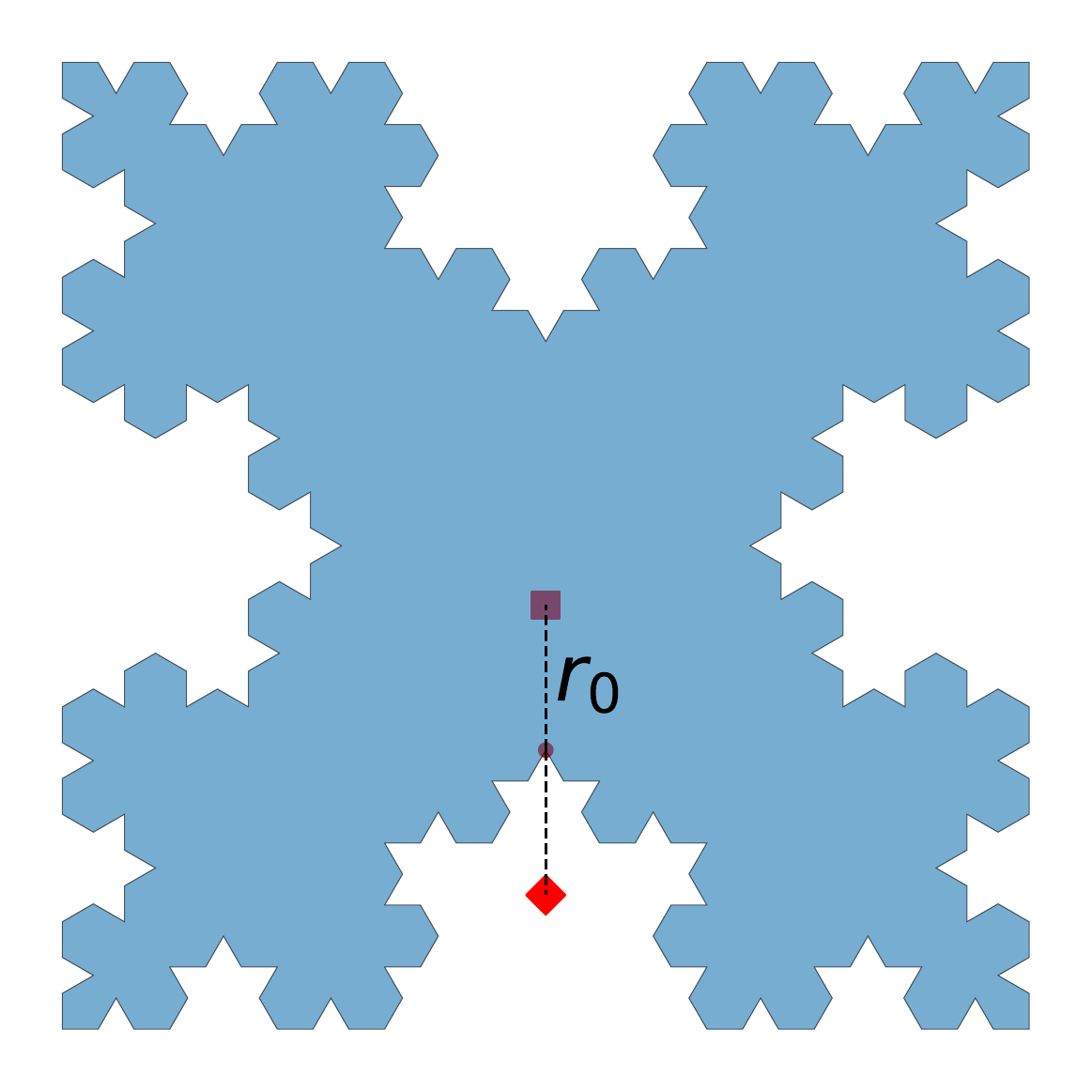}
  \caption{}
    \label{fig:Kochs_ext}
  \end{subfigure}
  \caption{
  \textbf{(a)} Schematic of an effective triangle based on a part of the Koch snowflake. 
The local environment of the Koch snowflake with angle $\Theta$ is shown by the shadowed region. Another isosceles triangle (transparent, enclosed by dash-dotted lines) shares the same vertex, with an effective angle $\Theta^\prime$ such that the areas of the two polygons are equal. 
\textbf{(b)} Inward-orientied Koch snowflake $\Omega^{4-}_{\frac{\pi}{3},3}$, constructed on the square. The starting point (red) is located on the vertical line close to the centering vertex with initial distance $r_0$. The diamond point presents an exterior starting position for Sec. \ref{sec:extcove}, whereas the square point presents an interior starting position for Sec. \ref{sec:intcave}. 
}
\end{figure}

For a wedge of angle $\Theta$, the persistence exponent is $\alpha = \frac{\pi}{2\Theta}$. 
In turn, the LPE for $\Omega_{\Theta,g}$ shows the mean value $\bar{\alpha}_{\Theta}$ that is smaller than $\frac{\pi}{2\Theta}$. 
The mean value $\bar{\alpha}_{\Theta}$ is obtained as the average: $\bar{\alpha}_{\Theta} = (\alpha^+_\Theta + \alpha^-_\Theta)/2$, where $\alpha^\pm_\Theta$ are the maximum and the minimum values of $\alpha_{\Theta,g}(t|r_0)$ in the log-periodic oscillation regime. 
In order to rationalize the dependence of $\bar{\alpha}_\Theta$ on $\Theta$, we employ a simple argument based on an \tit{effective} angle $\Theta^\prime$.
Consider the local environment of the Koch snowflake with the angle $\Theta$. We define an isosceles triangle with its vertex on the bottom, and two other vertices on the horizontal line of height $h_{\Theta,g-1}$ (Fig. \ref{fig:trimod}). 
We seek an \tit{effective} angle $\Theta^\prime$ such that the new isosceles triangle has the area $\mathcal{A}_{g-1}(\Theta^\prime)$ same as the local environment of the Koch snowflake. 
In other words, we suppose that the survival probability is primarily controlled by the area enclosed by the boundary, no matter the exact geometry. 
By a basic traigonometric analysis, the equivalent area requires $A_{\Theta,g-1} + 2 A_{\Theta,g} = \mathcal{A}_{g-1}(\Theta^\prime) = h_{\Theta,g-1} d$, where $d$ is shown on Fig. \ref{fig:trimod}.  
Then, $\Theta^\prime$ is solved by 
\begin{equation} 
\tan \llp \frac{\Theta^\prime}{2} \rrp = \frac{d}{h_{\Theta,g-1}} = \frac{\ell_{\Theta,g-1}}{h_{\Theta,g-1}} \sin\llp\frac{\Theta}{2}\rrp + \llp \frac{\ell_{\Theta,g}}{h_{\Theta,g-1}} \rrp^2 \sin\Theta  \,, 
\end{equation}
and thus $\bar\alpha^\mrm{eff}_\Theta$ is determined by
\begin{equation}
\bar\alpha^\mrm{eff}_\Theta = \frac{\pi}{2\Theta^\prime} = \frac{\pi}{4\arctan \llc \tan\frac{\Theta}{2} \lls 1 + \frac{1}{2 \llp 1 + \sin \frac{\Theta}{2} \rrp^2} \rrs \rrc} \,. 
\label{eq:meanalpha}
\end{equation}

Figure \ref{fig:amv} compares the numerical values of $\bar\alpha^\mrm{eff}_\Theta$ estimated from the curves $\alpha_{\Theta,g}(t|r_0)$ to Eq. (\ref{eq:meanalpha}). 
Despite the simplicity of this argument, we find acceptable agreement between numerical results and Eq. (\ref{eq:meanalpha}) over a broad range of angles. 
Moreover, there is no systematic bias on the relative error as a function of angle. 
For angles below $\frac{\pi}{3}$, we observe larger relative errors, while other angles show errors below $2\%$. 
Although this argument is oversimplified, it yields an estimate of the mean value of $\alpha_{\Theta,g}(t|r_0)$ at intermediate times.

Furthermore, Fig. \ref{fig:amv} shows how the amplitude, $\Delta\alpha_{\Theta} = \alpha^+_{\Theta} - \bar\alpha_{\Theta}$, 
changes with angle $\Theta$. 
For a better illustration, we plot the amplitude, multiplied by a factor of 10. 
The maximal amplitude is found at $\Theta = \frac{\pi}{3}$, perhaps due to the largest area of $A_{\Theta,1}$ (see Eq. (\ref{eq:Aareag})) among all angles $\Theta \in (0,\pi)$. 

\section{Extensions}
\label{Sec:exts}

In Sec. \ref{Sec:FPTinK}, we investigated log-periodic oscillations of LPEs for one family of Koch snowflakes with different angles $\Theta$. 
It arises naturally a question whether this behavior remains valid for other domains with similar fractal boundaries. 
In this section, 
we consider two other settings (Fig. \ref{fig:Kochs_ext}): 
(i) inward-oriented Koch snowflakes with particles diffusing outside the domain, and (ii) inward-oriented Koch snowflakes with particles diffusing inside the domain. 
We denote an inward-oriented domain constructetd on an $n$-regular polygon as $\Omega^{n-}_{\Theta,g}$, where the superscript ``$-$'' means that newly generated segments are inside the previous generation $\Omega^{n-}_{\Theta,g-1}$, and $\Theta,g$ are respectively the angle and the generation as before. 
The conventional outward-oriented Koch snowflakes generated on the equilateral triangle ($n = 3$) and studied in Sec. \ref{Sec:FPTinK} correspond to $\Omega^{3+}_{\Theta,g}$. 
As the boundaries of inward-oriented snowflakes $\Omega_{\Theta,g}^{3-}$ would be self-crossing for $\Theta < \pi/3$, we decided to replace the equilateral triangle ($n = 3$) by a square ($n = 4$) as the basic domain to construct inward-oriented domains $\Omega_{\Theta,g}^{4-}$, for which there is no self-crossing for any angle $\Theta \in (0,\pi)$. An example $\Omega_{\frac{\pi}{3},3}^{4-}$ is shown in Fig. \ref{fig:Kochs_ext}. 

\subsection{Exterior diffusion for inward-oriented Koch snowflakes}
\label{sec:extcove}

We employ again the GAFRW algorithm for Monte Carlo simulations in inward-oriented domains. 
The particle is initially located outside of the domain within the distance $r_0$ to the vertex, shown as the diamond in Fig. \ref{fig:Kochs_ext}. 
The only difference is the inclusion of an extra time threshold $T_\mrm{max} = 100$ to avoid too long trajectories. 
Once the diffusion time $t$ exceeds this threshold, we stop the current trajectory and then start the simulation for the next particle. 
There is no change in the analysis of FPTs.

\begin{figure}[t!]
\centering
\includegraphics[trim={0.2cm 0.2cm 1.0cm 1.0cm}, clip, width=0.48\linewidth]{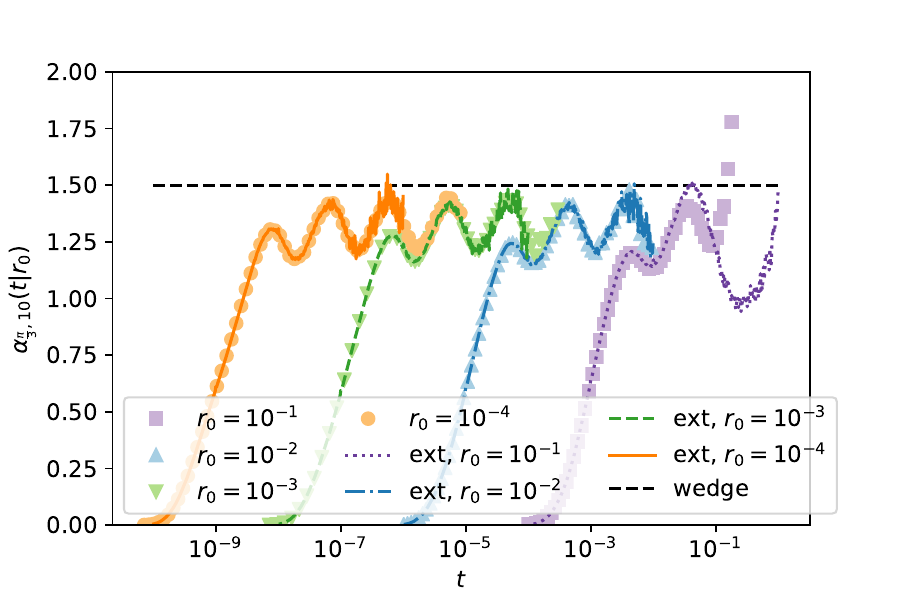}
\caption{
Comparison of the LPE $\alpha_{\frac{\pi}{3},10}(t|r_0)$ for outward-oriented Koch snowflake $\Omega^{3+}_{\frac{\pi}{3},10}$ (color dots) and inward-oriented domain $\Omega^{4-}_{\frac{\pi}{3},10}$ (color lines with ``ext'' in the legend) with $L=2$, $D=1$, and $N=10^8$ particles in Monte Carlo simulations. Distances are chosen as $r_0 = 10^{-4}, 10^{-3}, 10^{-2}, 10^{-1}$. 
The dashed horizontal line shows the persistence exponent 3/2 for the wedge of angle $\frac{\pi}{3}$. 
 }
\label{fig:extdomain}
\end{figure}

Figure \ref{fig:extdomain} compares the LPEs for the outward-oriented domain $\Omega_{\frac{\pi}{3},10}^{3+}$ and the inward-oriented domain  $\Omega_{\frac{\pi}{3},10}^{4-}$ with the same $L=2$, $g = 10$, and $\Theta = \frac{\pi}{3}$. 
We observe the almost identical behaviors for the two domains, as expected due to their identical local environment. 
The only exception appears for $r_0 = 10^{-1}$, where the violet dotted line does not increase in the same manner as the violet squares. 
As $h_{\frac{\pi}{3},2} \approx 0.192$ and $h_{\frac{\pi}{3},3} \approx 0.064$, we have $h_{\frac{\pi}{3},2} > r_0 > h_{\frac{\pi}{3},3}$ in this case. Hence, starting from the position at $r_0=10^{-1}$, particles are inclined to move to the center of the Koch snowflake $\Omega^{3+}_{\frac{\pi}{3},10}$ or to escape out of the square $\Omega^{4-}_{\Theta,0}$ at the time $t \approx 10^{-1}$. 
In the former case, diffusion in a confined domain implies $\alpha_{\frac{\pi}{3},10}(t|r_0) \propto t$ at long times, as observed in Fig. \ref{fig:extdomain}. In turn, the survival probability for a particle outside a bounded domain exhibits a logarithmic decay, $S(t|r_0) \propto \invs{\ln t}$ as $t \to \infty$ \cite{Redner}, and thus $\alpha(t|r_0) \propto \invs{(\ln t)^2} \to 0$. 
This asymptotic decay is very slow; moreover, in contrast to the interior diffusion, there is no characteristic timescale (like $L^2/D$), at which this decay is clearly established. In particular, we still observe oscillation-like behavior for the violet dotted curve for $t > 10^{-1}$, even though its amplitude is getting larger. 
We conclude that there is no difference between diffusion inside the outward-oriented domain and diffusion outside the inward-oriented domain in intermediate times; in contrast, distinctions emerge at long times.

\subsection{Interior diffusion for inward-oriented Koch snowflakes}
\label{sec:intcave}

\begin{figure}[t!]
  \centering
  \begin{subfigure}[b]{0.48\textwidth}
\includegraphics[trim={0.1cm 0.1cm 1.0cm 1.0cm}, clip, width=0.99\linewidth]{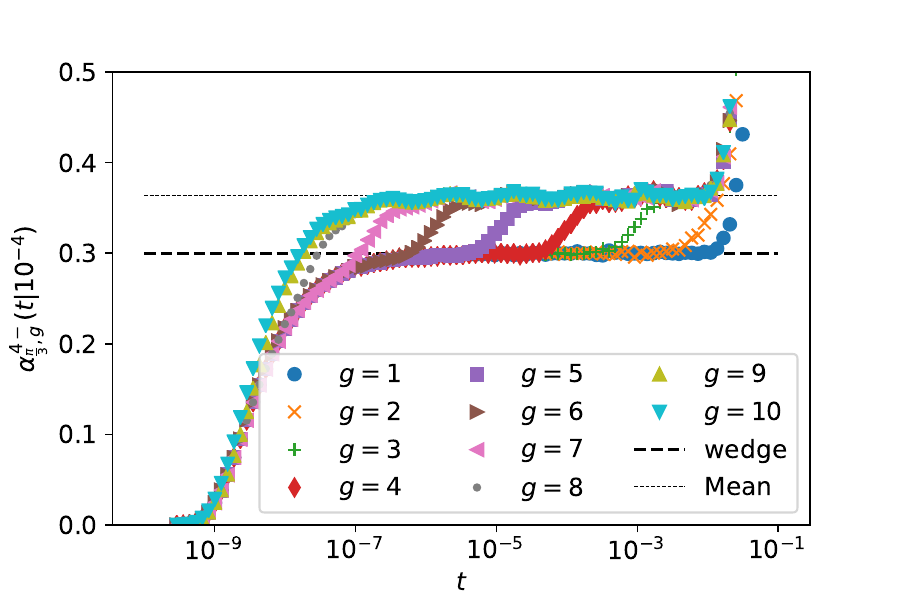}
    \caption{}
    \label{fig:extdomain_int-a}
  \end{subfigure}
  \hspace{1.0em} 
  \begin{subfigure}[b]{0.48\textwidth}
\includegraphics[trim={0.1cm 0.1cm 1.0cm 1.0cm}, clip, width=0.99\linewidth]{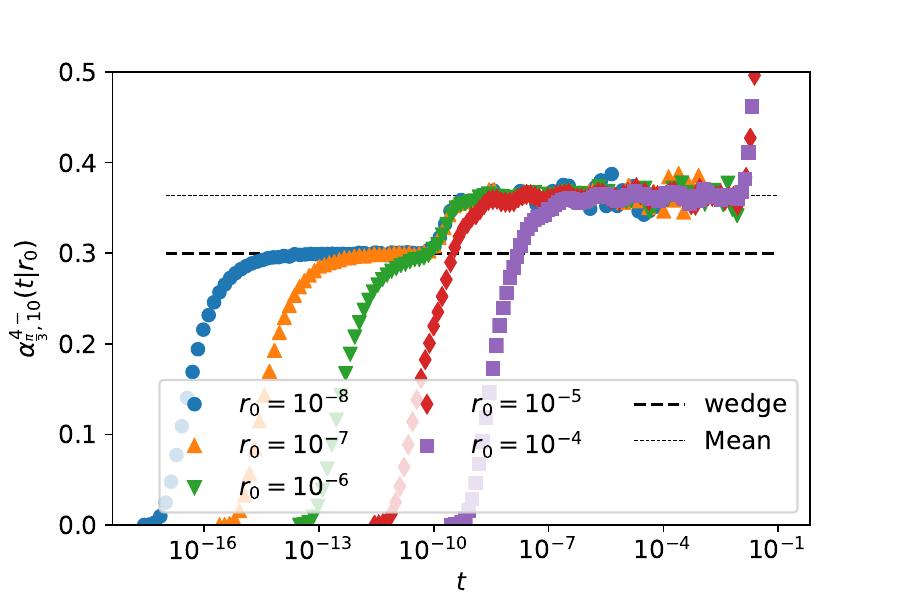}
  \caption{}
    \label{fig:extdomain_int-b}
  \end{subfigure}
  \caption{
  \textbf{(a)} Local persistence exponent $\alpha^{4-}_{\frac{\pi}{3},g}(t|10^{-4})$ for different generations $g$ (from 1 to 10), with $L=2$, and $D=1$, obtained by Monte Carlo simulations with $N = 10^8$ particles. 
The thick dashed horizontal line presents the persistence exponent $\frac{3}{10}$ for the wedge of angle $2\pi-\frac {\pi}{3} = \frac{5\pi}{3}$, while the thin dashed horizontal line approximates the mean value $0.36$ of log-periodic oscillations.  
\textbf{(b)} Local persistence exponent $\alpha^{4-}_{\frac{\pi}{3},10} (t|r_0)$ for different starting points, with $L = 2$ and $D = 1$, obtained by Monte Carlo simulations with $N = 10^8$ particles. 
The thick dashed horizontal line presents the persistence exponent $\frac{3}{10}$ for the wedge of angle $\frac{5\pi}{3}$, while the thin dashed horizontal line approximates the mean value $0.36$ of log-periodic oscillations.  
}
  \label{fig:extdomain_int}
\end{figure}

Another example concerns the starting point in the interior of the domain $\Omega^{4-}_{\pi/3,g}$, shown as the square in Fig. \ref{fig:Kochs_ext}. 
Figure \ref{fig:extdomain_int-a} compares LPEs for $\Omega^{4-}_{\Theta,g}$ as $g$ varies from 1 to 10, and uncovers a new feature as compared to the outward-oriented domains studied in Sec. \ref{Sec:FPTinK}. 
In fact, we observe two distinct regimes: (i) the wedge plateau, which reveals the effect of the local environment of the wedge of angle $2\pi-\frac {\pi}{3} = \frac{5\pi}{3}$, whose persistence exponent is $\frac{3}{10}$; and (ii) log-periodic oscillations around the mean value 0.36. Note that this mean value is below $\frac{\pi}{2[2\pi - (\pi+\Theta)/2]} = \frac{3}{8} = 0.375$, which is the persistence exponent of the wedge of angle $\frac{4\pi}{3}$. 
All curves exhibit transient growth at short and long times, as expected. 
For $g = 1, 2$, the particle only feels the local environment of a wedge at intermediate times, and LPEs stay on the wedge plateau before an increase at long times.
For other generations $g = 3, 4, 5$, apart from the wedge, the effect of self-similar boundary appears, so there exists a transition from the wedge plateau to the oscillation regime. 
For generations $g = 6, 7$, the wedge effect becomes weak as LPEs show no obvious stay on the wedge plateau. 
For generations $g = 8, 9, 10$, LPEs exceed the wedge plateau and advance towards the oscillation regime directly. 
Similar results were observed for other angles $\Theta$ (not shown). 

We argue that for a given parameter set $(\Theta, g, r_0)$, there exists a time threshold, before which the LPE $\alpha_{\Theta,g}(t|r_0)$ shows no transition from the wedge plateau to oscillations. 
To validate this statement, we also investigate the influence of the initial distance $r_0$ on LPEs. Figure \ref{fig:extdomain_int-b} presents the LPEs for different $r_0$ for the same generation $g = 10$, in which the transition time is about $t_t = 2\times10^{-10}$. 
For small $r_0 \le 10^{-7}$, LPEs reach the wedge plateau before $t_t$, and they follow the same transition. For the case $r_0 = 10^{-6}$, $\alpha_{\Theta,g}(t|r_0)$ arrives at the wedge plateau right after $t_t$, and it exhibits a direct transition towards log-periodic oscillations. 
For $r_0 \ge 10^{-5}$, the time instance for the LPE to reach the wedge plateau is after $t_t$, so we observe no transition but a direct growth towards log-oscillations. 
Finally, all curves grow at long times. 

As in the two previous cases (particles inside outward-oriented Koch snowflakes and particles outside inward-oriented domains), we observe transitions between the wedge plateau and oscillations, but in the opposite direction. 
Both settings furnish the same log-periods as approximated by Eq. (\ref{eq:validTlog}), which was also verified for other angles.

\section{Conclusion}
\label{Sec:con}

In this paper, we studied the first-passage time to absorbing self-similar boundaries of various Koch snowflakes. 
We investigated the local persistence exponent $\alpha_{\Theta,g}(t|r_0)$, focusing on its log-periodic oscillations at intermediate times. 
To explore this phenomenon, three different families of irregular shapes were inspected: 
(i) ordinary (outward-oriented) Koch snowflakes with an interior starting point; 
(ii) inward-oriented Koch snowflakes (Cesàro fractal) with an exterior starting point; 
(iii) inward-oriented Koch snowflakes with an interior starting point. 
For the first family, we observe that the domain with a larger angle (and thus a smaller fractal dimension $d_f$) produces the LPE with a lower mean value $\bar{\alpha}_{\Theta}$, a smaller amplitude of oscillations $\Delta \alpha_\Theta$, and a larger log-period $\Delta T_\Theta$. 
The second family allows us to investigate exterior diffusion near the self-similar boundary, which is locally identical to the first family; expectedly, we obtained very similar results for intermediate times, but significant deviations at long times. 
The third family illustrates the situation when the wedge plateau is lower than the mean value of log-periodic oscillations. 
Our extensive Monte Carlo simulations demonstrate that the LPE is a generic feature for domains with self-similar boundary, and the log-periodic oscillation is intrinsic to the associated fractal boundary. 
Moreover, we proposed and validated two simple approximations for the period $\Delta T_\Theta$ and the mean value $\bar{\alpha}_\Theta$ of log-oscillations. For instance, the period can be rationally explained by the area ratio between two adjacent generations, yielding $\Delta T_\Theta \propto 1/d_f$. 

The log-periodic behavior of the LPE reflects the self-similarity of the boundary $\partial \Omega_{\Theta,g}$, and the log-period can be understood as a suitable renormalization group in time. 
Therefore, there should exist the same behavior for the fractal boundary limit ($g \to \infty$). 
Since the log-periodic oscillation pattern does not depend on the generation $g$, (e.g., the mean values and log-periods of $\alpha_{\Theta,g}(t|r_0)$ are identical for different generations once the angle $\Theta$ and initial distance $r_0$ are fixed), we can attribute the log-periodic oscillation of $\alpha_{\Theta,g}$ as the approximation of $\alpha_{\Theta,\infty}$ at intermediate times. 
Moreover, the time of entering into the log-oscillation regime can be made arbitrarily small by diminishing $r_0$ if $g$ is infinite.

The presence of log-periodic oscillations in the behavior of the survival probability at intermediate time scales reveals that the diffusive dynamics near absorbing self-similar boundaries is much richer than earlier believed. A systematic analysis of the LPE in other disordered systems can improve our knowledge on chemical reactions in porous catalysts. In this light, further analysis for random fractals and three-dimensional systems would be informative. Moreover, as an immediate reaction upon the first arrival onto the boundary is an idealized picture, future studies of the survival probability near partially reactive boundaries present an important perspective. 
Another interesting direction for this study consists in relating log-periodic patterns to the geometrical structure of the eigenmodes of the Dirichlet Laplace operator in bounded domains, which may display self-similar features \cite{Sapoval91, Lapidus95, Even99, Daudert07, Banjai07, Grebenkov13}. Since the survival probability in such domains admits a spectral decomposition in terms of these eigenmodes, the observed behavior may reflect the underlying spectral geometry.


\section*{Acknowledgements}
D.S.G. acknowledges the Alexander von Humboldt Foundation for support within a Bessel Prize award. 
Y.Y acknowledges Adrien Chaigneau for fruitful discussions.

\end{document}